%% file: 00_main.tex
\pdfoutput=1

\documentclass[sigconf]{cidr-2026}
\usepackage[capitalize,nameinlink]{cleveref}
\usepackage{xspace}
\usepackage{minted}
\usepackage{graphicx}
\usepackage{subcaption}
\usepackage{enumitem}

\crefformat{section}{\S#2#1#3}
\crefformat{subsection}{\S#2#1#3}

\newcommand{\sys}{{\sc SPEAR}\xspace}

\setlist[itemize]{noitemsep, topsep=0pt, leftmargin=1em}

\begin{document}
\title{Making Prompts First-Class Citizens for Adaptive LLM Pipelines}

\author{U\u{g}ur \c{C}etintemel}
\affiliation{\institution{Brown University}}
\email{ugur\_cetintemel@brown.edu}

\author{Shu Chen}
\affiliation{\institution{Brown University}}
\email{shu\_chen@brown.edu}

\author{Alexander W. Lee}
\affiliation{\institution{Brown University}}
\email{alexander\_w\_lee@brown.edu}

\author{Deepti Raghavan}
\affiliation{\institution{Brown University}}
\email{deeptir@brown.edu}

\author{Duo Lu}
\affiliation{\institution{Brown University}}
\email{duo_lu@brown.edu}

\author{Andrew Crotty}
\affiliation{\institution{Northwestern University}}
\email{andrew.crotty@northwestern.edu}

\renewcommand{\shortauthors}{\c{C}etintemel et al.}

\input{01_abstract}
\maketitle

\input{01_intro}
\input{02_spear}
\input{03_opt}
\input{04_results}
\input{05_related}
\input{06_future}

\begin{acks}
This work is supported by a Brown Seed Grant and the National Science Foundation (NSF) Graduate Research Fellowship Program (GRFP) under grants 2439559 and 2040433.
Additional support was provided through Google's Research Scholar Program.
We also thank the members of the VectraFlow research team for their continued insights and constructive feedback about the project.
\end{acks}

\bibliographystyle{ACM-Reference-Format}
\bibliography{bib}

\end{document}

%% file: 01_abstract.tex
\begin{abstract}
Modern LLM pipelines increasingly resemble complex data-centric applications: they retrieve data, correct errors, call external tools, and coordinate interactions between agents. Yet, the central element controlling this entire process---the prompt---remains a brittle, opaque string that is entirely disconnected from the surrounding program logic. This disconnect fundamentally limits opportunities for reuse, optimization, and runtime adaptivity.

In this paper, we describe our vision and an initial design of \sys (\textit{Structured Prompt Execution and Adaptive Refinement}), a new approach to prompt management that treats prompts as first-class citizens in the execution model. Specifically, \sys enables: (1) structured prompt management, with prompts organized into versioned views to support introspection and reasoning about provenance; (2) adaptive prompt refinement, whereby prompts can evolve dynamically during execution based on runtime feedback; and (3) policy-driven control, a mechanism for the specification of automatic prompt refinement logic as \textit{when-then} rules.

By tackling the problem of runtime prompt refinement, \sys plays a complementary role in the vast ecosystem of existing prompt optimization frameworks and semantic query processing engines.
We describe a number of related optimization opportunities unlocked by the \sys model, and our preliminary results demonstrate the strong potential of this approach.
\end{abstract}

%% file: 01_intro.tex
\section{Introduction}
As LLMs are increasingly embedded in real-world systems, prompts have become the primary means of encoding user intent.
They steer data retrieval, direct output generation, recover from runtime errors, and coordinate tool use.
However, despite their centrality, prompts are often crafted manually in an unrigorous trial-and-error approach, and most LLM pipelines fail to appropriately version or track prompt evolution in a systematic fashion.
Even worse, they typically remain as static, opaque strings, completely detached from the broader program logic.

At the same time, these pipelines are rapidly evolving into complex data-centric applications that involve interactions with knowledge bases, conditional fallback, data validation, and multi-agent orchestration.
Several popular frameworks (e.g., LangChain~\cite{langchain}, DSPy~\cite{dspy}) allow developers to easily construct LLM processing pipelines, and semantic query engines~\cite{docetl,lotus,palimpzest,abacus} provide a declarative interface to LLM-augmented query processing.
Yet, in all of these approaches, prompt logic exists completely outside the system, a black box that stays entirely hidden to the optimizer and execution engine during runtime.

This paper proposes a radical shift in abstraction: treat prompts as structured data within the system.
We introduce our vision and an early design of \sys (\textit{Structured Prompt Execution and Adaptive Refinement}), which makes three core contributions:
\begin{itemize}
\item \textbf{Structured Prompt Management.} Prompts are treated as rich, structured data objects and organized into named views, which promotes modularity and reusability of program logic. Prompt views are also versioned so that developers can introspect them and track their evolution over time. We define a flexible prompt algebra for creating and refining prompts within an application, which facilitates end-to-end optimization of entire pipelines.
\item \textbf{Adaptive Prompt Refinement.} In \sys, prompts are not static templates tuned offline but rather part of the system state that evolves dynamically at runtime. Refinements are operators that modify prompt logic, potentially based on signals available only during execution. \sys allows for a range of refinement modes from fully manual to completely automated, giving developers fine-grained control over the whole refinement process.
\item \textbf{Policy-Driven Control:} \sys also introduces \textit{when-then} policies that govern how and under what circumstances to trigger automatic refinements. These policies can apply refinements either before execution based on operator inputs (e.g., selecting a task-specific prompt view) or adaptively after execution by considering runtime feedback (e.g., result confidence, latency threshold, shared token budget). \sys's policy layer unifies static planning and runtime adaptivity, establishing a principled mechanism for application-level refinement control.
\end{itemize}

Although DSPy~\cite{dspy} aims to replace manual prompt engineering with automated optimization, it operates primarily offline and offers limited runtime refinement capabilities~\cite{dspy-assert,gepa}.
As such, \sys addresses the complementary dimension of runtime prompt refinement, allowing adaptivity of prompt logic based on real-time signals and feedback.
This perspective surfaces completely new opportunities for end-to-end optimization, and our preliminary results demonstrate the promise of this approach.

%% file: 02_spear.tex
\section{SPEAR}
\label{sec:sys}
This section provides a high-level overview of our approach to prompt management, called \sys (\textit{Structured Prompt Execution and Adaptive Refinement}).
We begin with a running example to highlight the pain points of current approaches and then introduce each of the key concepts underlying \sys's design.

\subsection{Running Example}
\label{sec:example}

Consider an application that uses LLMs to help a clinician review a patient's medical history while forming a diagnosis and developing a corresponding treatment plan.
One pipeline might focus on extracting and reasoning about clinical notes mentioning medications like enoxaparin, a commonly prescribed anticoagulant.

To implement this pipeline, a developer would likely begin with a straightforward prompt: \textit{``Summarize the patient's medication history and highlight any use of enoxaparin.''}
After observing that initial outputs are inconsistent, with some omitting dosage and others missing timing or administration information, the developer might then attempt to improve the results manually by adding to the prompt: \textit{``Be specific about dosage and indicate whether enoxaparin was administered within the last 48 hours.''}
As the pipeline becomes more complicated, the developer will need to begin maintaining prompt variants to apply in specific scenarios (e.g., additional information about drug interactions for inpatients) while also handling low-confidence responses and other exceptions.
Although a variety of existing frameworks (e.g., LangChain~\cite{langchain}, DSPy~\cite{dspy}) attempt to alleviate some of this burden, the developer is ultimately left with the manual and labor-intensive task of updating the pipeline via brittle ad hoc prompt edits.

\sys adopts a fundamentally different perspective: prompts are rich, structured data objects that evolve over time and can be directly manipulated based on runtime signals during execution.
As such, \sys blurs the line between the program (i.e., prompt logic) and the data upon which it operates, allowing developers to write pipelines that can directly reason over and optimize their own prompt strategies.

\subsection{Structured Prompt Management}
As mentioned, \sys's key insight is to elevate prompts to first-class citizens by treating them just like any other data in the system.
In \sys, prompts are not simple strings but rather structured data objects that, in addition to the actual prompt text, store other metadata like runtime statistics and version history as the prompt evolves over time.
In the following, we describe how prompts are defined, used, and finally compiled to an executable prompt algebra.

\smallskip\noindent\textbf{Prompt Views.}
Like views in SQL, \sys represents each prompt as a logical view that encapsulates a reusable piece of program logic expressed in natural language.
For instance, in the example from \cref{sec:example}, the developer would define an \texttt{enoxaparin\_summary} view that could then be invoked anywhere in the application requiring that information.

Also similar to SQL views, prompt views are composable, such that they are definable in terms of other named views.
Developers can therefore decompose complex prompt logic into small, modular building blocks.
For example, suppose that \texttt{enoxaparin\_summary} required information about drug interactions for inpatients with ongoing treatment; in this case, the developer could refine the prompt to conditionally use the \texttt{check\_interactions} view based on patient status.
We cover prompt refinement in \cref{sec:refinement}.

Lastly, \sys supports prompt view parameterization, allowing developers to define a prompt template that is invoked with specific values at runtime.
Continuing to build on the running example, the developer could replace \texttt{enoxaparin\_summary} with a generic \texttt{medication\_summary} prompt view that takes the medication name as a parameter, since the application will likely need to provide similar summaries for many different medications.

Unlike other approaches, prompt views offer key advantages from a developer perspective in terms of code modularity, reusability, and overall maintainability.
This abstraction also presents a range of optimization opportunities, which we discuss in \cref{sec:opt}.

\sys maintains the set of all prompt views $P$ as a key-value mapping of named views to prompt objects.
Likewise, the context store $C$ contains all runtime data over which prompt views operate, including inputs, outputs, and intermediate results.
Prompt views may additionally reference or update the runtime metadata $M$, a collection of control signals and diagnostic information generated during execution (e.g., shared token budget, latency threshold).
Taken together, the prompt views $P$, context store $C$, and runtime metadata $M$ comprise the system state.

\smallskip\noindent\textbf{Prompt Algebra.}
At the heart of \sys is an executable prompt algebra that operates on the system state in a structured and principled way.
\sys's algebra is closed under composition, in that every operator must consume and produce a triple of the form $(P, C, M)$.
This design naturally makes the set of operators both flexible and easily extensible.

\sys has several built-in core operators, the most important of which is \texttt{GENERATE} (\texttt{GEN}).
Using the specified LLM, \texttt{GEN} applies a prompt view to any referenced data in $C$ and stores the result back into $C$.
Other core operators include \texttt{RETRIEVE} (\texttt{RET}), which fetches data from an external data source (e.g., database, web search, API) into $C$, and \texttt{SWITCH}, which encodes conditional logic for pipeline control flow.
In \cref{sec:summary}, we show how the end-to-end pipeline we have built up as part of our running example compiles to a plan with these three operators---including additional prompt refinement logic, which we discuss next.

\subsection{Adaptive Prompt Refinement}
\label{sec:refinement}
So far, we have mainly described how developers build pipelines that update the context store $C$ and runtime metadata $M$.
However, since prompts are treated exactly the same as all other data in \sys, operators can also directly introspect, reason about, and modify prompt logic.
Importantly, this capability extends to dynamic refinement at runtime that can adapt to real-time execution feedback, which sets \sys apart from existing frameworks that focus mostly on offline prompt optimization.

More formally, a refinement is any operator that updates the set of prompt views $P$ by modifying existing prompts or composing entirely new ones, which results in a \texttt{REFINE} (\texttt{REF}) edge in the corresponding prompt algebra plan.
These refinements might update an existing prompt by appending examples, injecting hints, or rewriting confusing instructions.
Next, we explain the different refinement modes in \sys that let the developer balance control and automation.
We then introduce a policy-based mechanism that allows the developer to specify how \sys should apply refinements automatically.

\begin{figure*}[t]
  \centering
  \begin{subfigure}[t]{\textwidth}
  \centering
  \begin{minipage}{0.91\textwidth}
  %\begin{minted}[frame=single,fontsize=\footnotesize]{python}
  \begin{minted}[fontsize=\footnotesize]{python}
  medication_summary = "Summarize the patient's medication history and highlight any use of {medication}."
  medication_summary += " Be specific about dosage and indicate whether {medication} was administered within the last 48 hours."
  medication_summary += " " + check_interactions(medication) if is_inpatient else ""
  output = medication_summary(clinical_notes, medication='enoxaparin', is_inpatient=True)
  \end{minted}
  \end{minipage}
  \end{subfigure}
  \begin{subfigure}[t]{0.9\textwidth}
  \centering
  %\frame{
  \includegraphics[width=0.9\textwidth,trim={0mm 13mm 0mm 12mm},clip]{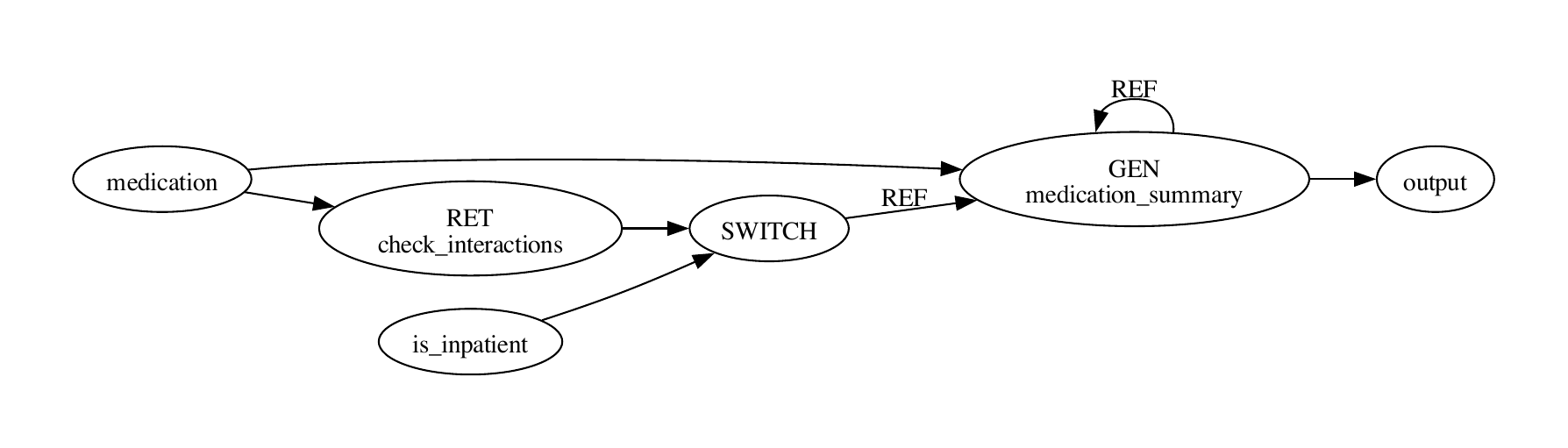}
  %}
  % Answer: [trim={left bottom right top},clip]
  \end{subfigure}
  \vspace{-3mm}
  \caption{Final implementation of the running example from \cref{sec:example}, including developer code (top) and compiled plan (bottom).}
  \vspace{-1mm}
  \label{fig:summary}
\end{figure*}

\smallskip\noindent\textbf{Refinement Modes.}
Depending on the degree of oversight that the developer desires, \sys supports a range of refinement modes that fall into three broad categories:

\begin{itemize}
\item\textbf{Manual.} This mode is useful when the developer requires low-level control over the refinement (e.g., enforcing a specific output format) or must inject relevant domain knowledge into the prompt. We have already shown several manual refinements applied to the running example from \cref{sec:example}, including the addition of a request for specific details in the output summary and updating the prompt to accept the medication as a parameter. In all of these cases, the developer explicitly modified the prompt text.

\item\textbf{Assisted.} This mode offers a more declarative refinement option, where the developer only needs to provide a high-level description of intent. For example, an assisted alternative to the original manual refinement of the enoxaparin prompt shown in \cref{sec:example} might look like: \textit{``Be sure to include details about recent enoxaparin administration.''} In our prompt algebra, this assisted refinement would compile to a \texttt{GEN} operator that consumes the original prompt and produces a new version, which would ideally look similar to the manual refinement.

\item\textbf{Automatic.} This mode passively monitors the system state at runtime and triggers a refinement based on a user-specified condition. As with assisted refinements, \sys will compile the refinement to a prompt algebra plan for execution, all without any developer intervention. We will revisit automatic refinements, including concrete examples and how to specify them, when we introduce policy-driven control.
\end{itemize}

In practice, these modes often coexist and can even be combined depending on risk tolerance or system maturity.
For example, an application might begin with manual refinement during early development, transition to assisted prompting based on observed patterns, and ultimately move toward automatic handling for scale and responsiveness.
Conversely, in deployed settings, pipelines may default to automatic refinement, escalate to assisted repair when needed, and fall back to manual oversight in ambiguous or high-risk cases.
\sys does not prescribe a particular refinement mode but rather provides the flexibility to choose different modes based on application needs and developer preferences.

\smallskip\noindent\textbf{Policy-Driven Control.}
As noted, both the manual and assisted refinement modes rely on active developer involvement to determine when to update a prompt.
In other words, these types of refinements are applied statically at compile time in response to an explicit command.
Automatic refinements, though, are triggered during runtime in response to the current system state.

In \sys, refinement policies allow the developer to specify the circumstances in which to apply an automatic refinement to a prompt.
Specifically, we define a policy as a simple \textit{when-then} rule of the form: \texttt{WHEN condition THEN refinement}.
As the name suggests, the developer can include one or more refinement actions in \texttt{refinement}.
Signals referenced by \texttt{condition}, which actually triggers the refinement, may originate from the context store $C$, runtime metadata $M$, or even a prompt's own metadata (e.g., repeated failed retries, degradation in accuracy).
Depending on what system state \texttt{condition} references, the application of \texttt{refinement} can occur either before or after the evaluation of the prompt algebra plan for the corresponding view:

\begin{itemize}
\item\textbf{Input Condition (Before).} When \texttt{condition} depends only on inputs to an operator (e.g., parameters, materialized intermediate results), \sys can apply all refinements before the operator is even evaluated. For instance, the \texttt{check\_interactions} refinement from the running example can be determined solely based on a patient's status before any execution takes place.

\item\textbf{Output Condition (After).} On the other hand, if \texttt{condition} must examine the output of an operator, then \sys can only apply refinements after executing that operator. For example, if the output from \texttt{medication\_summary} does not include all of the necessary information as determined by an LLM judge, the policy could automatically issue a retry or fall back to a different LLM/prompt combination.
\end{itemize}

While the system state captured in $(P, C, M)$ constitutes \sys's data plane, refinement policies can be thought of as the control plane that determines how and when pipelines automatically evolve.
However, they should not be thought of as a fixed set of hard-coded rules.
In fact, refinement policies can even be implemented as prompts, meaning that they compile directly into \sys's prompt algebra and can themselves be adaptively refined.
For example, policies that trigger infrequently or rarely improve output quality can be deprioritized or pruned, whereas those that consistently provide significant benefits can be promoted.
As such, policy-driven refinement control could lay the foundation to build a fully tunable and self-improving control plane with \sys.

\subsection{Putting It All Together}
\label{sec:summary}
Returning to the example from \cref{sec:example}, \cref{fig:summary} shows the complete, end-to-end implementation of the LLM pipeline for generating medication summaries from clinical notes.
The developer code (top) begins with a definition of the parameterized \texttt{medication\_summary}, followed by the manual refinement concerning required content for the output.
The next line shows the refinement policy that conditionally applies the \texttt{check\_interactions} view for inpatients.
Until now, the developer has defined several operators but not actually invoked any of them; note that \sys evaluates pipelines lazily, so they exist only as unrendered templates at this point.

On the next line, the developer calls \texttt{medication\_summary} with the arguments \texttt{medication=`enoxaparin'}, \texttt{is\_inpatient=True}, and a specific patient's \texttt{clinical\_notes} as context.
This invocation causes \sys to compile \texttt{medication\_summary} into the executable prompt algebra plan shown in \cref{fig:summary} (bottom), which contains two named inputs, one named output, and three \sys operators.
In the plan, \texttt{GEN} will call an LLM to produce output for the main prompt view, \texttt{medication\_summary}.
\texttt{SWITCH} implements refinement policy logic that will conditionally apply the previously defined \texttt{check\_interactions} view, instantiated here as a \texttt{RET} operation from a database of known drug interactions.

The two \texttt{REF} edges in the plan denote each of the refinements applied to the \texttt{medication\_summary} view.
As mentioned, some refinements can be resolved entirely at compile time, such as the \texttt{REF} that appended additional instructions to the original prompt to ensure that the summary would contain specific details.
On the other hand, \sys must apply the second \texttt{REF}, which depends on an input value, conditionally at runtime.

The developer is now ready to either execute the compiled plan as a standalone pipeline or embed it into an orchestration framework like LangChain for deployment.

%% file: 03_opt.tex
\section{Optimization Opportunities}
\label{sec:opt}
The combination of structured prompt management and adaptive prompt refinement in \sys unlocks a range of optimization opportunities.
In this section, we highlight examples of such optimizations that a system adopting \sys's approach could implement.

\smallskip\noindent\textbf{Prefix Caching \& Reuse.}
Reusing attention states is a common technique to accelerate prompt evaluation~\cite{pope}.
Since large parts of the prompt may remain unchanged across successive invocations in many pipelines, \sys could identify stable portions of a prompt and cache them for reuse in the future.
When making only a small refinement to a prompt (e.g., adding an example), \sys will usually append the delta to the end of the prompt view rather than rebuilding the entire prompt from scratch.
This incremental construction enables straightforward token-level reuse with LLMs that support a KV cache~\cite{vllm} or FlashAttention~\cite{flashattention}, reducing latency and compute costs.
To generalize this idea, \sys could employ a structured prompt cache~\cite{promptcache} that indexes prompt views and their rendered forms, facilitating efficient reuse across retries or invocations with different arguments.

\smallskip\noindent\textbf{Input Batching.}
When the same prompt view is applied repeatedly to many different input items (e.g., invoking \texttt{medication\_summary} for a variety of different medications), \sys could instead execute them as a single batched \texttt{GEN} operation, allowing the LLM to process all inputs in one call.
This approach is common in semantic query processing engines~\cite{docetl,palimpzest,abacus}, where identical query logic is applied concurrently across batches of inputs to maximize throughput by filling as much of the LLM context window as possible.
However, input batching also introduces risks like prompt crowding, degraded accuracy due to cross-input interference, and challenges in isolating errors.
\sys will need to mitigate these challenges by carefully tuning the batch size to balance performance gains with output quality based on workload characteristics, model behavior, and downstream quality requirements, as well as providing a fallback option to reprocess inputs that cause issues.

\smallskip\noindent\textbf{Prompt Batching.}
Similar to input batching, the main idea is to exploit redundancy in the structure of prompts.
This situation occurs frequently in \sys, since prompt views are constructed from modular and reusable building blocks.
Prompt batching submits groups of similar prompts in a single LLM call, amortizing dispatch overhead while enabling parallel inference without significant risk of cross-prompt interference.
When prompts share a common prefix, they may additionally benefit from the previously described prefix caching technique, further improving efficiency.

\smallskip\noindent\textbf{Operator Fusion.}
\sys can support runtime operator fusion via refinements, updating operators to merge them into a single execution unit that can improve efficiency and reduce the size of intermediate result storage.
For tightly coupled operators, such as prompt views that operate over different parts of the same input, operator fusion is a natural optimization.
However, unlike in traditional settings where operator fusion is almost always beneficial, fusing prompts can actually result in worse accuracy or higher latency.
As such, striking the right balance when deciding which or how many operators to fuse in \sys presents an interesting opportunity for further exploration.

%% file: 04_results.tex
\section{Preliminary Results}
\label{sec:results}

%Inference costs (USD per 1M tokens) for the models used in our evaluation. We use the commercial pricing of Gemma-3-4B on Together AI~\cite{togetherai} as our anchor. The 1B model cost is estimated based on competitive serving rates for similar 1B-parameter models (e.g., Llama-3.2-1B~\cite{requesty_llama1b}). Due to the lack of publicly available serving costs for the 12B and 27B variants at the time of writing, we infer these costs by extrapolating proportionally based on parameter count relative to the 4B anchor.

\noindent\textbf{Experimental Setup.}
We conducted all experiments on a server with a single NVIDIA Quadro RTX 8000 GPU (48 GB GDDR6) running a local Ollama instance for model hosting.
Our results include an evaluation of four different model sizes from Google's Gemma 3~\cite{gemma3} model family: 1B, 4B, 12B, and 27B.
We estimated approximate inference costs for each model from publicly available data, which we report in \cref{tab:model_pricing}.

To evaluate \sys, we selected EHRNoteQA~\cite{ehrnoteqa}, a question-answering benchmark based on clinical notes from real-world electronic health records.
The benchmark, which contains 962 unique multiple-choice questions with options \textit{A--E}, mimics the types of pipelines described in the running example from \cref{sec:example}.
For an answer to be marked correct, the prediction must be a single letter (after stripping whitespace) that exactly matches (including capitalization) the ground truth label.
Each experiment was repeated three times on the same randomly chosen 200 samples, with mean values reported for accuracy and total cost.

% \todo{Breakdown of individual refinement strategies, including manual (p1, p2), automatic/agentic (p3), and runtime (p4-p6).}
%We evaluate a progressive series of prompt refinement strategies, organized into three categories:
%\textit{Manual Refinements (p1-p2):} These represent offline, human-designed prompt improvements.  while 
%\textit{Agentic Refinement (p3):} 
% This evidence-first ordering is designed to improve the model's grounding in the provided clinical notes before attempting to answer.
%%\textbf{p1} extends the baseline by adding explicit output format constraints (``Respond with exactly one letter and no further explanation''),
%In the post-execution policy, the prompt is augmented with an additional option ``F. No other choice is clearly supported by the evidence.'' Models are instructed not to guess. If a model returns F (uncertain), the request cascades to a larger model in the pool. If all models return F, the largest model is queried using the p3 prompt as a fallback. For p5, we evaluate three different model pool configurations: [1B, 3B, 12B, 27B], [3B, 12B, 27B], and [12B, 27B]. For p6 with router-based routing, we test each model in the pool as the router. 
%This likely aids the model's attention mechanism by grounding the thinking process in the source text immediately. 

\begin{figure*}[t]
    \centering
    \includegraphics[width=\textwidth]{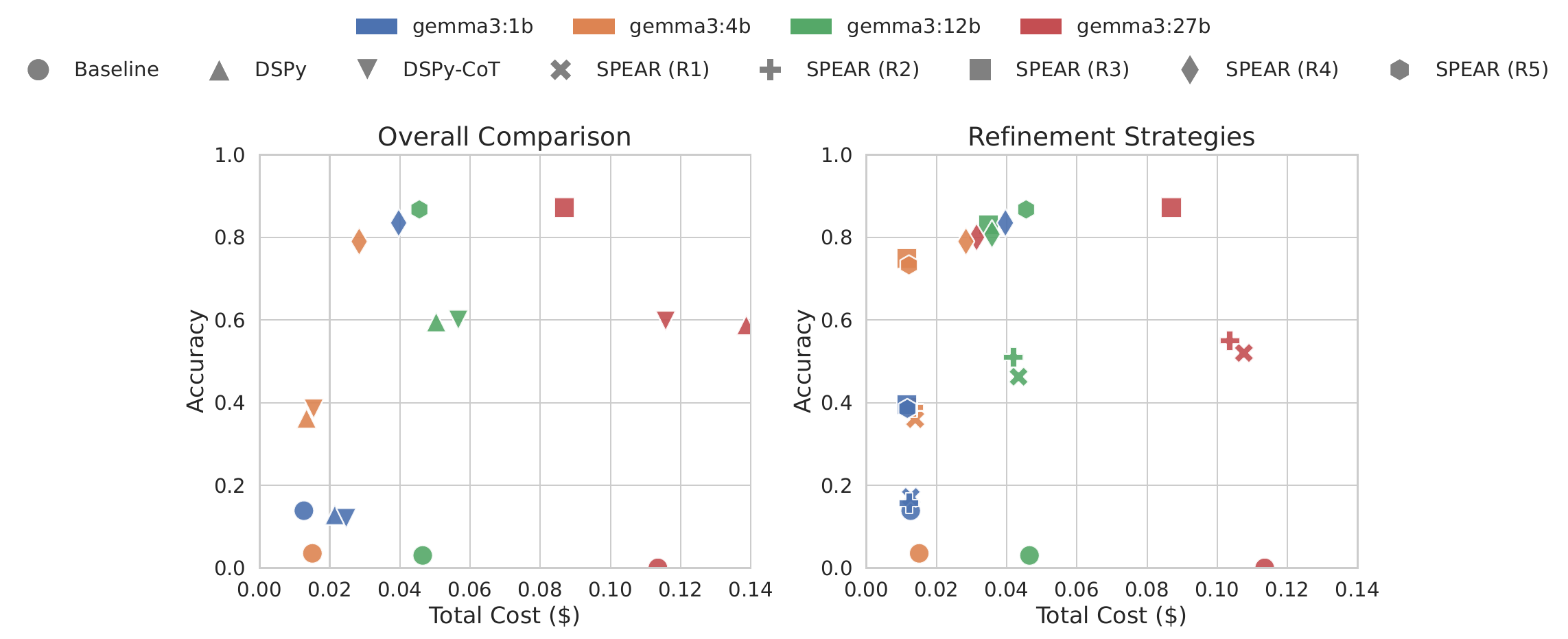}
    \caption{Results of our evaluation, including overall comparison (left) and refinement strategy breakdown (right).}
    \label{fig:ehrnoteqa}
\end{figure*}

\smallskip\noindent\textbf{Comparison Points.}
We compare \sys against: (1) \textit{Baseline}, a minimal prompt with only basic instructions for answering JSON-formatted multiple-choice questions; (2) \textit{DSPy}, using the \texttt{Predict} module to manually extract relevant fields from the JSON object; and (3) \textit{DSPy-CoT}, which uses DSPy's built-in chain-of-thought module.
For \sys, we include the following refinements:
\begin{itemize}
\item\textbf{R1 (manual, static).}
Extends the baseline by adding explicit instructions about output formatting.

\item\textbf{R2 (assisted, static).}
Adjusts the input structure of \textit{R1} by extracting the question, answer choices, and clinical notes from the JSON object into clearly delineated sections with explicit labels, as suggested offline by the 27B model.

\item\textbf{R3 (assisted, static).}
Slightly rearranges the prompt components of \textit{R2} to place the clinical evidence first, followed by the question and answer choices, and ending with the output constraints, again as suggested offline by the 27B model.

\item\textbf{R4 (automatic, input condition).}
Based on the routing model's assessment of the question's difficulty, routes the question to an appropriate model and then uses \textit{R3}.

\item\textbf{R5 (automatic, output condition).}
Adds an additional answer option to \textit{R3} that allows the model to select $F$ when uncertain, then cascades to the next larger model until reaching the largest, which uses \textit{R3}.
\end{itemize}

\begin{table}[t]
    \centering
    \caption{Estimated inference costs for each model.}
    %\vspace{-1}
    \label{tab:model_pricing}
    \begin{tabular}{lcc}
        \toprule
        \textbf{Model} & \textbf{Input (\$/1M)} & \textbf{Output (\$/1M)} \\
        \midrule
        gemma3:1B  & 0.02 & 0.02 \\
        gemma3:4B  & 0.02 & 0.04 \\
        gemma3:12B & 0.06 & 0.12 \\
        gemma3:27B & 0.15 & 0.30 \\
        \bottomrule
    \end{tabular}
\end{table}

\smallskip\noindent\textbf{Discussion.}
\cref{fig:ehrnoteqa} shows the results, with the overall comparison on the left (including the best \sys strategy for each model) and a detailed breakdown of \sys refinement strategies on the right.
We observe several key trends:
\begin{itemize}
\item\textbf{Comparison to DSPy.}
\textit{DSPy} and \textit{DSPy-CoT} provide a substantial improvement over the baseline, but they are consistently outperformed by many of the \sys refinements.
Despite attempting to enforce the output constraint via a \texttt{Literal[...]} field, DSPy often struggles to strictly constrain the generation of chatty models in long-context scenarios, leading to parsing errors similar to the baseline.
Furthermore, DSPy's internal prompt optimization process is opaque, making it difficult to diagnose and debug specific failure modes compared to \sys's explicit, transparent approach to prompt refinement, which allows for much more targeted fixes.

\item\textbf{Impact of Prompt Structure.}
We observe a clear progression in performance through our refinement stages.
The baseline prompt performs poorly mainly because most models fail to output a single letter as an answer without explicit instructions to do so.
They often generate verbose reasoning chains that complicate answer extraction and lead to frequent formatting errors, a problem that is even more noticeable with larger models.
\textit{R3} exhibits the most dramatic leap in performance across every model, which highlights the value of our structured and transparent refinement process; unlike opaque optimization methods, we can explicitly identify and apply interpretable changes like prompt view component reordering.

\item\textbf{Efficiency of Automatic Refinements (R4, R5).}
The automatic refinement strategies successfully optimize the trade-off between cost and accuracy by exploiting the heterogeneity of question difficulty in the benchmark.
In particular, \textit{R4} demonstrates consistently strong performance by dynamically matching model capacity to query complexity.
Notably, using the smallest 1B model as a router yields an impressive accuracy of 83.5\%, matching the performance of the standalone 12B model but with the flexibility to call larger models only when needed.
This result confirms that a significant portion of the dataset consists of relatively easy questions that do not require the larger, more expensive models to get a correct answer.
For \textit{R5}, a cascade starting with the 12B model achieves 86.7\% accuracy, comparable to the standalone 27B model (87.3\%) but at nearly half the cost (\$0.046 vs. \$0.087).
By only escalating queries when the smaller model explicitly signals uncertainty about its output, this automatic refinement ensures that more expensive model calls are reserved solely for the most challenging questions.
\end{itemize}

Although techniques like \textit{R4} and \textit{R5} are not entirely novel, \sys enables developers to implement these complex optimization strategies in a simple and modular fashion.
By treating them as composable refinement primitives, \sys lays the groundwork for reusing them as building blocks in more sophisticated optimization strategies.
Furthermore, this structured approach can allow \sys to adaptively refine the routing logic or output uncertainty thresholds based on accumulated runtime feedback.

%% file: 05_related.tex
\section{Related Work}
\label{sec:related}

\noindent\textbf{Prompt Programming Frameworks.}
Numerous recent frameworks support the creation of complex LLM pipelines and agentic workflows through chaining, orchestration, and dynamic feedback loops (e.g., LangChain~\cite{langchain}).
However, they commonly treat prompts as static strings, with little support for structured prompt management, introspection, or optimization.
In contrast, \sys promotes prompts to first-class entities with an executable prompt algebra that supports reuse and adaptive runtime refinement.

\smallskip\noindent\textbf{Manual Prompt Refinement.}
Prompt engineering techniques, such as adding explicit instructions, incorporating domain-specific examples, or restructuring task descriptions, have been shown to significantly improve output accuracy, consistency, and alignment with task goals across a wide range of applications~\cite{sivarajkumar,Wang}.
For instance, approaches like MedPrompt~\cite{medprompt} demonstrate that structured prompts can significantly improve accuracy and relevance in clinical LLM pipelines.
\sys's refinements aim to support the straightforward application of similar techniques to prompt logic in an easily controllable fashion.

\smallskip\noindent\textbf{Automated Prompt Optimization.}
Frameworks like DSPy~\cite{dspy} allow developers to provide high-level specifications in order to automate the prompt optimization process.
DSPy Assertions~\cite{dspy-assert} allow for the specification of constraints on model outputs and enable assertion-driven self-refinement both offline and at runtime, whereas techniques like GEPA~\cite{gepa} refine prompts offline by reflecting on execution trajectories to iteratively propose and select improved prompt candidates.
SPADE~\cite{spade} also maintains prompt histories from user edits for the purpose of synthesizing assertions before deployment.
Despite some runtime capabilities, these systems mainly target offline prompt optimization, positioning \sys as a complementary approach that can help dynamically refine prompts based on feedback during execution.

\smallskip\noindent{\textbf{Semantic Query Engines.}}
Recent LLM-augmented semantic data processing systems~\cite{docetl,lotus,palimpzest,abacus} offer strong data-level semantics, but their control over LLM behavior is limited; that is, prompt logic is still inlined or statically defined and cannot adapt dynamically during execution.
\sys fills a different role and could even be paired with these systems to handle runtime prompt refinement.

%% file: 06_future.tex
\section{Conclusion \& Future Work}
This paper described \sys, our new approach to prompt management based on the core tenet that prompts should be elevated from simple strings to first-class citizens in LLM pipelines.
In doing so, \sys unifies structured prompt management, adaptive refinement, and policy-driven control into a coherent execution model.
Our approach enables richer forms of prompt introspection, principled reuse, and dynamic optimization, all of which are difficult or impossible in existing frameworks.
Early results suggest that this perspective opens the door to a broad class of optimizations that can substantially improve robustness, efficiency, and developer productivity.
In future work, we plan to continue to develop \sys and further explore the many fruitful optimization opportunities we highlighted throughout the paper.

%% file: bib.bib
@inproceedings{dspy,
  author       = {Omar Khattab and
                  Arnav Singhvi and
                  Paridhi Maheshwari and
                  Zhiyuan Zhang and
                  Keshav Santhanam and
                  Sri Vardhamanan and
                  Saiful Haq and
                  Ashutosh Sharma and
                  Thomas T. Joshi and
                  Hanna Moazam and
                  Heather Miller and
                  Matei Zaharia and
                  Christopher Potts},
  title        = {{DSPy: Compiling Declarative Language Model Calls into State-of-the-Art Pipelines}},
  booktitle    = {{ICLR}},
  year         = {2024},
}

@inproceedings{ehrnoteqa,
  author       = {Sunjun Kweon and
                  Jiyoun Kim and
                  Heeyoung Kwak and
                  Dongchul Cha and
                  Hangyul Yoon and
                  Kwang Kim and
                  Jeewon Yang and
                  Seunghyun Won and
                  Edward Choi},
  title        = {{EHRNoteQA: An LLM Benchmark for Real-World Clinical Practice Using Discharge Summaries}},
  booktitle    = {{NeurIPS}},
  year         = {2024},
}

@article{medprompt,
  author       = {Harsha Nori and
                  Yin Tat Lee and
                  Sheng Zhang and
                  Dean Carignan and
                  Richard Edgar and
                  Nicol{\`{o}} Fusi and
                  Nicholas King and
                  Jonathan Larson and
                  Yuanzhi Li and
                  Weishung Liu and
                  Renqian Luo and
                  Scott Mayer McKinney and
                  Robert Osazuwa Ness and
                  Hoifung Poon and
                  Tao Qin and
                  Naoto Usuyama and
                  Christopher M. White and
                  Eric Horvitz},
  title        = {{Can Generalist Foundation Models Outcompete Special-Purpose Tuning? Case Study in Medicine}},
  journal      = {{CoRR}},
  volume       = {abs/2311.16452},
  year         = {2023},
}

@article{gemma3,
  author       = {Gemma Team},
  title        = {{Gemma 3 Technical Report}},
  journal      = {{CoRR}},
  volume       = {abs/2503.19786},
  year         = {2025},
}

@inproceedings{palimpzest,
    title={{Palimpzest: Optimizing AI-Powered Analytics with Declarative Query Processing}},
    author={Liu, Chunwei and Russo, Matthew and Cafarella, Michael and Cao, Lei and Chen, Peter Baile and Chen, Zui and Franklin, Michael and Kraska, Tim and Madden, Samuel and Shahout, Rana and Vitagliano, Gerardo},
    booktitle = {{CIDR}},
  year         = {2025},
}

@article{Wang,
  author       = {Li Wang and
                  Xi Chen and
                  Xiangwen Deng and
                  Hao Wen and
                  Mingke You and
                  Weizhi Liu and
                  Qi Li and
                  Jian Li},
  title        = {{Prompt engineering in consistency and reliability with the evidence-based guideline for LLMs}},
  journal      = {{npj Digit. Medicine}},
  volume       = {7},
  number       = {1},
  year         = {2024},
}

@article{sivarajkumar,
  author       = {Sonish Sivarajkumar and
                  Mark Kelley and
                  Alyssa Samolyk{-}Mazzanti and
                  Shyam Visweswaran and
                  Yanshan Wang},
  title        = {{An Empirical Evaluation of Prompting Strategies for Large Language Models in Zero-Shot Clinical Natural Language Processing}},
  journal      = {{CoRR}},
  volume       = {abs/2309.08008},
  year         = {2023},
}

@inproceedings{flashattention,
  author       = {Tri Dao and
                  Daniel Y. Fu and
                  Stefano Ermon and
                  Atri Rudra and
                  Christopher R{\'{e}}},
  title        = {{FlashAttention: Fast and Memory-Efficient Exact Attention with IO-Awareness}},
  booktitle    = {{NeurIPS}},
  year         = {2022},
}

@article{gepa,
  author       = {Lakshya A. Agrawal and
                  Shangyin Tan and
                  Dilara Soylu and
                  Noah Ziems and
                  Rishi Khare and
                  Krista Opsahl{-}Ong and
                  Arnav Singhvi and
                  Herumb Shandilya and
                  Michael J. Ryan and
                  Meng Jiang and
                  Christopher Potts and
                  Koushik Sen and
                  Alexandros G. Dimakis and
                  Ion Stoica and
                  Daniel Klein and
                  Matei Zaharia and
                  Omar Khattab},
  title        = {{GEPA: Reflective Prompt Evolution Can Outperform Reinforcement Learning}},
  journal      = {{CoRR}},
  volume       = {abs/2507.19457},
  year         = {2025},
}

@inproceedings{promptcache,
  author       = {In Gim and
                  Guojun Chen and
                  Seung{-}Seob Lee and
                  Nikhil Sarda and
                  Anurag Khandelwal and
                  Lin Zhong},
  title        = {{Prompt Cache: Modular Attention Reuse for Low-Latency Inference}},
  booktitle    = {{MLSys}},
  year         = {2024},
}

@article{lotus,
  author       = {Liana Patel and
                  Siddharth Jha and
                  Melissa Z. Pan and
                  Harshit Gupta and
                  Parth Asawa and
                  Carlos Guestrin and
                  Matei Zaharia},
  title        = {{Semantic Operators and Their Optimization: Towards AI-Based Data Analytics with Accuracy Guarantees in LOTUS}},
  journal      = {{PVLDB}},
  volume       = {18},
  number       = {11},
  pages        = {4171--4184},
  year         = {2025},
}

@inproceedings{pope,
  author       = {Reiner Pope and
                  Sholto Douglas and
                  Aakanksha Chowdhery and
                  Jacob Devlin and
                  James Bradbury and
                  Jonathan Heek and
                  Kefan Xiao and
                  Shivani Agrawal and
                  Jeff Dean},
  title        = {{Efficiently Scaling Transformer Inference}},
  booktitle    = {{MLSys}},
  year         = {2023},
}

@article{abacus,
  author       = {Matthew Russo and
                  Sivaprasad Sudhir and
                  Gerardo Vitagliano and
                  Chunwei Liu and
                  Tim Kraska and
                  Samuel Madden and
                  Michael J. Cafarella},
  title        = {{Abacus: A Cost-Based Optimizer for Semantic Operator Systems}},
  journal      = {{CoRR}},
  volume       = {abs/2505.14661},
  year         = {2025},
}

@article{docetl,
  author       = {Shreya Shankar and
                  Tristan Chambers and
                  Tarak Shah and
                  Aditya G. Parameswaran and
                  Eugene Wu},
  title        = {{DocETL: Agentic Query Rewriting and Evaluation for Complex Document Processing}},
  journal      = {{PVLDB}},
  volume       = {18},
  number       = {9},
  pages        = {3035--3048},
  year         = {2025},
}

@article{spade,
  author       = {Shreya Shankar and
                  Haotian Li and
                  Parth Asawa and
                  Madelon Hulsebos and
                  Yiming Lin and
                  J. D. Zamfirscu{-}Pereira and
                  Harrison Chase and
                  Will Fu{-}Hinthorn and
                  Aditya G. Parameswaran and
                  Eugene Wu},
  title        = {{SPADE: Synthesizing Data Quality Assertions for Large Language Model Pipelines}},
  journal      = {{PVLDB}},
  volume       = {17},
  number       = {12},
  pages        = {4173--4186},
  year         = {2024},
}

@article{dspy-assert,
  author       = {Arnav Singhvi and
                  Manish Shetty and
                  Shangyin Tan and
                  Christopher Potts and
                  Koushik Sen and
                  Matei Zaharia and
                  Omar Khattab},
  title        = {{DSPy Assertions: Computational Constraints for Self-Refining Language Model Pipelines}},
  journal      = {{CoRR}},
  volume       = {abs/2312.13382},
  year         = {2023},
}

@misc{vllm,
  title     = {{Automatic Prefix Caching}},
  howpublished = {\url{https://docs.vllm.ai/en/stable/features/automatic_prefix_caching/}},
  year         = {2025},
}

@misc{langchain,
  title     = {{LangChain}},
  howpublished = {\url{https://www.langchain.com/}},
  year         = {2025},
}
